\documentclass[prl,twocolumn,showpacs]{revtex4}
\usepackage{epsfig}
\usepackage{graphicx}
\usepackage{mathrsfs}
\usepackage{amssymb}
\usepackage{color}
\usepackage[toc,page]{appendix}

\newcommand{\bcen}{\begin{center}}
\newcommand{\ecen}{\end{center}}
\newcommand{\btab}{\begin{tabular}}
\newcommand{\etab}{\end{tabular}}
\newcommand{\bdes}{\begin{description}}
\newcommand{\edes}{\end{description}}

\newcommand{\beq}{\begin{equation}}
\newcommand{\eeq}{\end{equation}}
\newcommand{\bea}{\begin{eqnarray}}
\newcommand{\eea}{\end{eqnarray}}

\newcommand{\bary}{\begin{array}}
\newcommand{\eary}{\end{array}}
\newcommand{\benum}{\begin{enumerate}}
\newcommand{\eenum}{\end{enumerate}}
\newcommand{\bitem}{\begin{itemize}}
\newcommand{\eitem}{\end{itemize}}

\newcommand{\bk} { {\mathbf k} }

\newcommand{\br} { \mbox{\boldmath $r$}}

\newcommand{\bQ} { \mbox{\boldmath $Q$}}

\newcommand{\bS} { \mbox{\boldmath $S$}}

\newcommand{\mean}[1]{\langle #1 \rangle}
\newcommand{\bra}[1]{{\langle #1 |}}
\newcommand{\ket}[1]{| #1 \rangle}

\newcommand{\Fig}[1]{Fig.~\ref{#1}}

\newcommand{\Neel}{N\'eel}

\begin{document}

\title{Competition between antiferromagnetism and superconductivity, electron-hole doping asymmetry and ``Fermi Surface" topology in cuprates}
\author{Sandeep~Pathak$^{1}$, Vijay B.~Shenoy$^{2,1}$, Mohit Randeria$^{3}$, and Nandini Trivedi$^{3}$}
\affiliation{$^{1}$ Materials Research Centre, Indian Institute of Science, Bangalore 560 012, India\\
$^{2}$ Centre for Condensed Matter Theory, Department of Physics, Indian Institute of Science, Bangalore 560 012, India\\
$^{3}$ Department of Physics, The Ohio State University, 191 W. Woodruff Avenue, Columbus, OH 43210 }
%% \author{Vijay B.~Shenoy}
%% \email[]{shenoy@physics.iisc.ernet.in}
%% %\homepage[]{Your web page}
%% %\altaffiliation{}
%% \affiliation{Department of Physics, Indian Institute of Science, Bangalore 560 012, India}
%% \author{Nandini~Trivedi}
%% \email[]{trivedi@mps.ohio-state.edu}
%% \affiliation{Department of Physics, The Ohio State University, Columbus}
%% \author{Mohit~Randeria}
%% \email[]{randeria@mps.ohio-state.edu}
%% \affiliation{Department of Physics, The Ohio State University, Columbus}

\begin{abstract}
We investigate the asymmetry between electron and hole doping in a 2D Mott insulator,
and the resulting competition between antiferromagnetism (AF) and
d-wave superconductivity (SC), using variational Monte Carlo for projected wave functions.
We find that key features of the $T=0$ phase
diagram, such as critical doping for SC-AF coexistence and the maximum value of the SC order
parameter, are determined by a single parameter $\eta$ which characterises the topology of the ``Fermi
surface" at half filling defined by the bare tight-binding parameters.
Our results give insight into why AF wins for electron doping, while SC is dominant on the hole doped side. We also suggest using band structure engineering to control the $\eta$ parameter for enhancing SC.
\end{abstract}


\maketitle

Ever since their discovery, cuprates continue to
pose some of the most challenging theoretical puzzles~\cite{NatPhy2006} in condensed matter physics.   
The problem is dominated by strong
electronic correlations~\cite{Anderson1987,Lee2006,Orenstein2000,Schrieffer2007} and the $tJ$-model and its variants
are believed to contain the essential physics of high
$T_c$ superconductivity. In this Letter we address the following questions:
What controls the electron-hole asymmetry in cuprates? Why does antiferromagnetism dominate on the electron doped side, and superconductivity on the hole doped side?
How can we understand the empirical correlation between electronic structure parameters -- the range of the in-plane hopping -- and superconductivity, pointed out by Pavarini {\it et al.}~\cite{Pavarini2001}?
In particular, can we get some insight into the all important question of what material parameters control the optimal SC transition temperature 
$T_c^{\rm max}$? 

\noindent {\bf Model:} 
The minimal model that allows for an understanding of the material
dependencies of the cuprate pheonomenology is the $t$-$J$ model with extended hopping:
${\cal H} =  -P \sum_{i,j, \sigma}  
t_{i,j}(c^\dagger_{i\sigma} c_{j \sigma}  + \mbox{h.~c.} ) P
+  J\sum_{\mean{i,j}} (\bS_i \cdot \bS_j - {n}_i {n}_j /4)$
where $c_{i \sigma}$ is the electron operator at site $i$ with spin $\sigma$, $n_{i\sigma}=c^\dagger_{i\sigma}c_{i\sigma}$
is the density, with $n_i=\sum_\sigma n_{i\sigma}$,
$\bS_i = \frac{1}{2} c^\dagger_{i\alpha}\vec{\sigma}_{\alpha\beta} c_{i\beta} $ is the spin
at site $i$ ($\sigma$'s are the Pauli matrices), and $J$ is
the antiferromagnetic exchange between nearest neighbors $\mean{i,j}$. 
The projection operator $P=\prod_i \left(1-n_{i\uparrow}n_{i\downarrow}\right)$ implements the 
``no double occupancy'' constraint. 

The bare dispersion has the form
$\epsilon(\bk) = -2 t (\cos{k_x} + \cos{k_y}) + 4 t' \cos{k_x} \cos{k_y} -2 t''(\cos{2k_x} + \cos{2k_y})$
where $t$, $t'$ and $t''$ are the nearest, second and third-neighbor hoppings respectively.
The importance of $t'$ and $t''$ is suggested both by ARPES experiments~\cite{Campuzano2004,Damascelli2003}
and electronic structure calculations~\cite{Andersen1995}. 
With the sign convention above, $t,t'$ and $t''$ are all positive for the hole doped case. 
To model the electron-doped case, we make a standard particle-hole transformation 
$\tilde{c_i}=(-1)^ic_i$ and $\tilde{c}^{\dagger}_i=(-1)^ic^{\dagger}_i$
in ${\cal H}$. Thus for the electron-doped case we again obtain ${\cal H}$ with $\tilde{t}=t$, $\tilde{t'}=-t'$, $\tilde{t''}=-t''$.  

\noindent {\bf Variational Wave function:}
We choose a variational ground state wave function for an $N$-particle system that includes both AF and SC order:
\bea
\ket{\Psi_{0}} = 
P [\sum_{ij} \varphi({\bf r}_i-{\bf r}_j) c^\dagger_{i \uparrow} c^\dagger_{j \downarrow} ]^{N/2} \ket{0}  \label{HO}
\eea
The form of $\varphi$ in the unprojected wave function is motivated by a saddle point analysis of ${\cal H}$.
For a nonzero Neel amplitude $m_N$, we get two spin density wave (SDW) bands ($\alpha = 1,2$) 
$
E_{(1,2)}(\bk) =[\xi(\bk) + \xi(\bk+\bQ)]/2
\pm (1/2) \sqrt{ (\xi(\bk) - \xi(\bk+\bQ))^2 + 16 J^2 m_N^2 }
$
where $\bQ=(\pi,\pi)$ and $\xi(\bk)=-2 t (\cos{k_x} + \cos{k_y}) + 4 t^\prime_{\rm {var}} \cos{k_x} \cos{k_y}
-2 t^{\prime\prime}_{\rm {var}}(\cos{2k_x} + \cos{2k_y}) -\mu_{\rm {var}}$, 
with $\bk$ in the {\it reduced} Brillouin zone. 
The $d$-wave pairing field  
$\Delta(\bk) = J \Delta(\cos{k_x} - \cos{k_y})$ gives rise to SC in the two SDW bands, with BCS coherence factors
\bea
\frac{v_{\bk\alpha}}{u_{\bk\alpha}} = (-1)^{\alpha-1} \frac{\Delta(\bk)}{E_{\alpha}(\bk) + \sqrt{E_{\alpha}^2(\bk) + \Delta^2(\bk)}}.
\eea 
The internal pair wave function $\varphi({\bf r}_i-{\bf r}_j)$ is given by the Fourier transforming
 ${v_{\bk\alpha}}/{u_{\bk\alpha}}$,
and summing over the two SDW bands $\alpha$. 

\begin{figure*}
%\centerline{\epsfxsize=\widefig \epsfbox{phdiag.eps}}
\centerline{\epsfxsize=4.5in \epsfbox{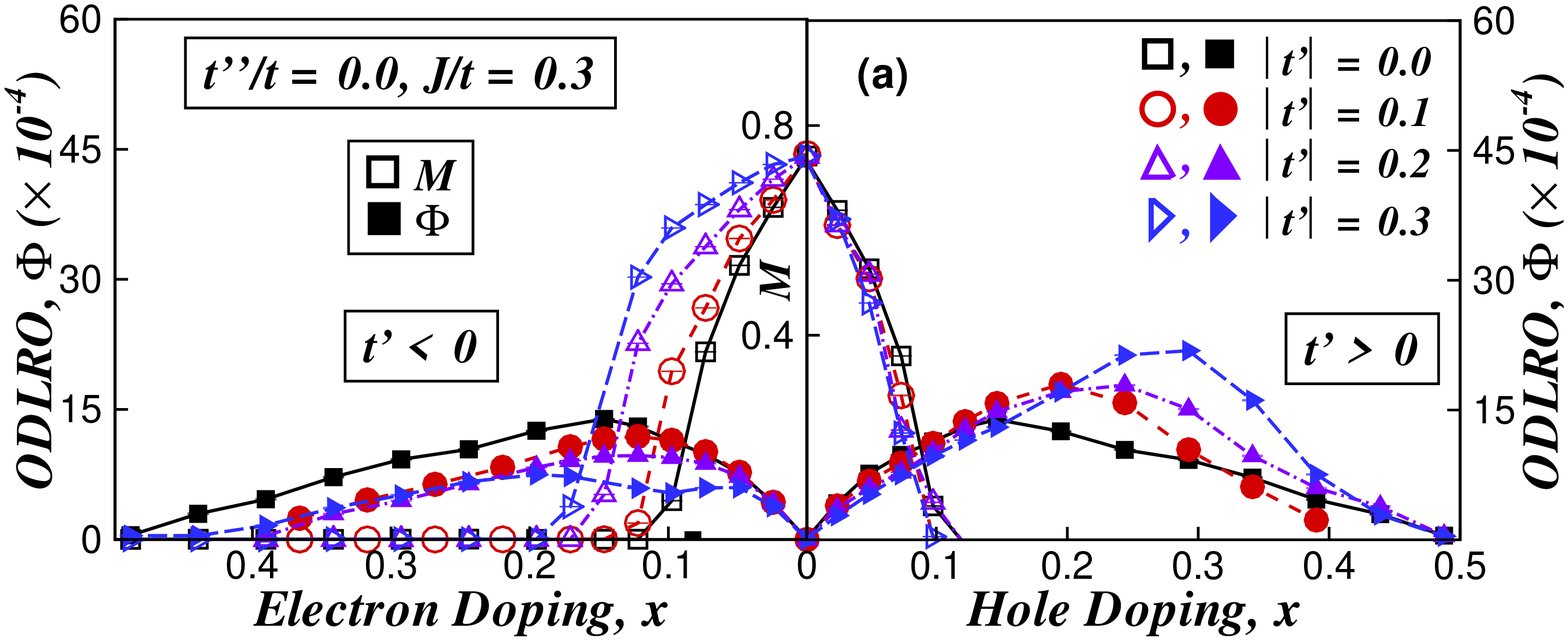}}  
\centerline{\epsfxsize=1.25in \epsfclipon \epsfbox[87 046 351 270]{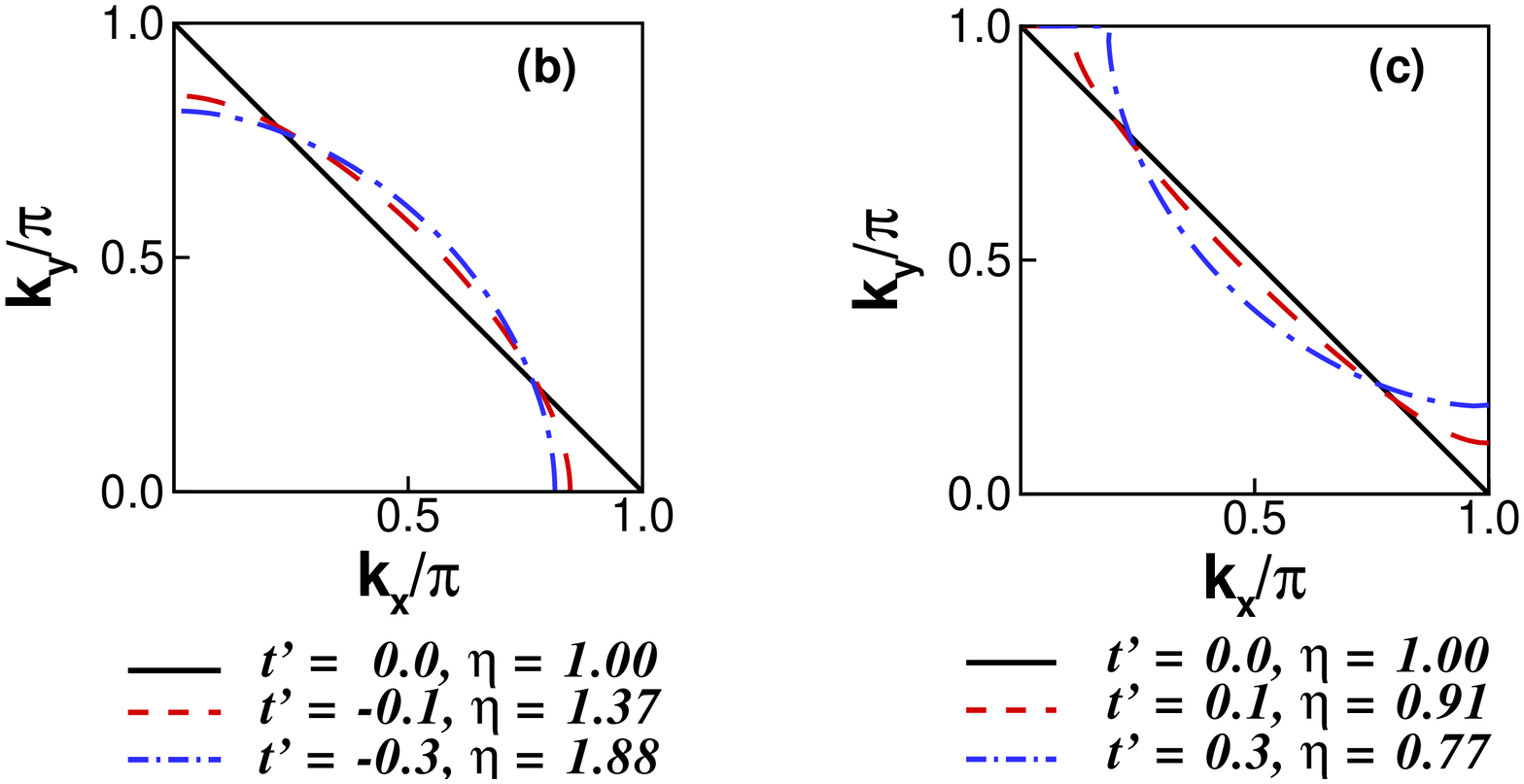}~~
\epsfxsize=3.75in \epsfclipon \epsfbox[35 271 751 561]{fsconvVsconc.eps}
\epsfxsize=1.25in \epsfclipon \epsfbox[481 046 745 270]{fsconvVsconc.eps}
}
%  \centerline{\epsfxsize=2.0in \epsfbox{fstpm.eps}\epsfxsize=2.0in \epsfbox{fstpp.eps}}
  \caption{{\small (color online) (a) Phase Diagram of extended $t$-$J$ model with $t''=0$ and $t'$ indicated in the inset, showing
\Neel \ magnetization $M$ and the SC order parameter $\Phi$ as a function of both hole- and electron-doping. 
(b),(c) Shapes of the bare Fermi surface at half filling.}}
 \label{PhDiag}
\end{figure*}

The five variational parameters in $\ket{\Psi_{0}}$ are the \Neel \ amplitude $m_N$,
the $d$-wave gap $\Delta$, the (Hartree-shifted) chemical potential $\mu_{\rm {var}}$, 
and $t^{\prime}_{\rm {var}}$ and $t^{\prime\prime}_{\rm {var}}$ renormalized by Fock shifts.
Their optimum values are determined by
minimizing the ground state energy 
$E = \bra{\Psi_0} {\cal H}\ket{\Psi_0}/ 
\langle{\Psi_0}\vert {\Psi_0}\rangle$ 
calculated using variational Monte Carlo (VMC), which exactly implements the 
projection $P$.
We have developed a fast conjugate gradient algorithm (details will be published elsewhere) which evaluates 
derivatives of the energy and efficiently determines the optimized variational
parameters.

%\begin{figure*}
%%\centerline{\epsfxsize=\widefig \epsfbox{phdiag.eps}}
%\centerline{\epsfxsize=4.5in \epsfbox{phdiag.eps}}  
%\centerline{\epsfxsize=1.25in \epsfclipon \epsfbox[87 046 351 270]{fsconvVsconc.eps}~~
%\epsfxsize=3.75in \epsfclipon \epsfbox[35 271 751 561]{fsconvVsconc.eps}
%\epsfxsize=1.25in \epsfclipon \epsfbox[481 046 745 270]{fsconvVsconc.eps}
%}
%
%%  \centerline{\epsfxsize=2.0in \epsfbox{fstpm.eps}\epsfxsize=2.0in \epsfbox{fstpp.eps}}
%  \caption{{\small (color online) (a) Phase Diagram of extended $t$-$J$ model with $t''=0$ and $t'$ indicated in the
%inset, showing
%\Neel \ magnetization $M$ and the SC order parameter $\Phi$ as a function of both hole- and electron-doping. 
%(b),(c) Shapes of the bare Fermi surface at half filling.}}
% \label{PhDiag}
%\end{figure*}

\noindent {\bf Results:}
The $T=0$ phase diagram is determined by computing the SC and AF order parameters for the optimized ground state as a function of doping~\cite{lattice}. The SC order parameter~\cite{Paramekanti2004} $\Phi= \lim_{|\br -\br'| \longrightarrow \infty} F_{\alpha,\alpha} (\br-\br')$ is obtained from the long range behavior of the 
correlation function
$ F_{\alpha,\beta}(\br-\br') = \mean{B^{\dagger}_{\br\alpha}B_{\br'\beta}}$ 
where, $B^{\dagger}_{\br\alpha} = \frac{1}{2}(c^{\dagger}_{\br\uparrow}c^{\dagger}_{\br+\hat{\alpha}\downarrow}-c^{\dagger}_{\br\downarrow}c^{\dagger}_{\br+\hat{\alpha}\uparrow})$ creates a singlet on the bond $(\br,\br+\hat{\alpha})$, $\alpha = x, y$. 
The AF order parameter
$M = ({2}/{N})\mean{\sum_{i\;\epsilon A} S^{z}_i - \sum_{i\;\epsilon B} S^{z}_i}$
is the difference of the magnetization
on the $A$ and $B$ sublattices.

In \Fig{PhDiag} we see the following phases: an AF Mott insulator at half-filling; coexistence of SC and AF 
for small (electron/hole) doping; a $d$-wave SC at higher doping; and a Fermi liquid for sufficiently large doping.
The phase diagram shows marked 
electron-hole asymmetry for $t'\ne 0$. As the next-neighbor hopping $|t'|$ increases, SC is
enhanced on the hole doped side, while AF is
stabilized on the electron doped side. These
results are consistent with earlier VMC~\cite{Chen1990,Giamarchi1991,Lee1997,Himeda1999,Lee2003,Shih2004,Shih_prl2004,Paramekanti2004,Anderson2004}
and dynamical mean field studies~\cite{Kotliar2001,Tremblay2005}.
  
Our new findings are that, on the hole doped side,
$|t'|$ does not affect AF, and in particular the critical doping
$x_{AF}$, beyond which AF vanishes, is insensitive to the value of $|t'|$. On the
electron doped side, SC is slightly weakened and the peak
value of the SC order parameter $\Phi_{max}$ falls with increasing $|t'|$.

Upon adding a second neighbor hopping $t''$ we find the following general trends:
(a) On the hole doped side, an increase in $t''$ leads to an
increase of superconducting correlations. Interestingly, AF is
relatively unaffected, and $x_{AF}$ is quite insensitive to $t''$.
(b) On the other hand, on the electron doped side, AF is enhanced
with increasing $|t''|$, and superconductivity is essentially
unaffected. We have also performed calculations using a simple
renormalized mean-field theory~\cite{Zhang1988,Anderson2004}, and the
qualitative phase diagram is in agreement with the VMC
results.

At very low doping, our results differ from experiments due to the neglect 
of long range Coulomb and disorder effects. Once the local superfluid density,
becomes sufficiently small near half-filling due to projection \cite{Paramekanti2004},
long-wavelength, quantum phase fluctuations, neglected in our approach,
drive SC to zero at a finite doping~\cite{rajdeep}. 
We also do not consider the effects of other competing orders (stripes or charge ordering). 
For large doping, where the pairing $\Delta\rightarrow 0$, finite size effects 
become large because of a growing correlation length. This leads to
an overestimate~\cite{Spanu2007} of the range over which SC survives, but
does not qualitatively affect our conclusions.

\noindent {\bf ``Fermi surface" topology:} 
Is there a simple way to understand our results?
To begin with, we emphasize that the dependence of the phase diagram on the bare dispersion is {\it not} controlled by the
van Hove singularity (vHS) in the bare density of states. Even in the particle-hole symmetric case $t'=t''=0$, where
the vHS is precisely at the center of the band, the calculated SC order parameter $\Phi(x)$ does not peak at $x=0$ but rather at an `optimal' doping away from zero. This `optimal' doping is determined by the interplay between the growth of the pairing amplitude 
$\Delta$ with underdoping and the suppression of phase coherence by strong correlations as $x\rightarrow 0$.
Irrespective of the values of $t'$ and $t''$, the renormalized bandwidth is of order $(xt+J)$, while the scale of the pairing is also 
of order $J$. Thus we are not in a weak coupling BCS regime where all the action is in the immediate vicinity of the chemical potential. In fact, the entire band participates in pairing, 
and the proximity of the vHS to the chemical potential is not the dominant factor in determining the phase diagram. 

We next define a quantity $\eta$ 
that characterizes the topology of the ``Fermi
surface" at half filling, defined by the bare tight-binding parameters.
We show that \emph{this single parameter $\eta$ controls 
the dependence of the phase diagram on} $t'$ \emph{and} $t''$. 
In particular, key features such as the
critical doping $x_{AF}$, the maximum value of the SC
order parameter $\Phi_{max}$, and optimal doping, are all determined by $\eta$.

\begin{figure}
\includegraphics[width=1.5in,angle=0]{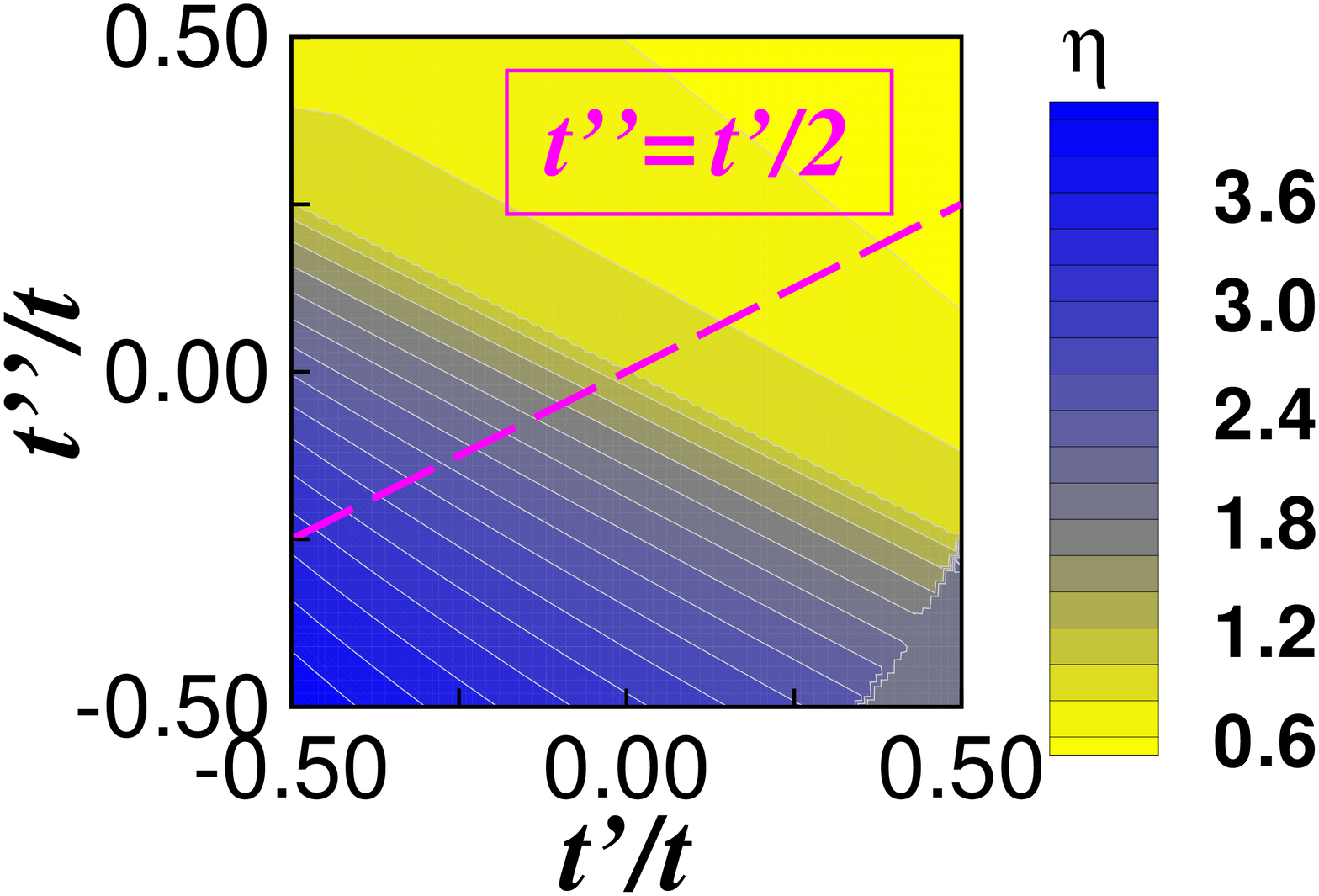}
\includegraphics[width=1.4in,angle=0]{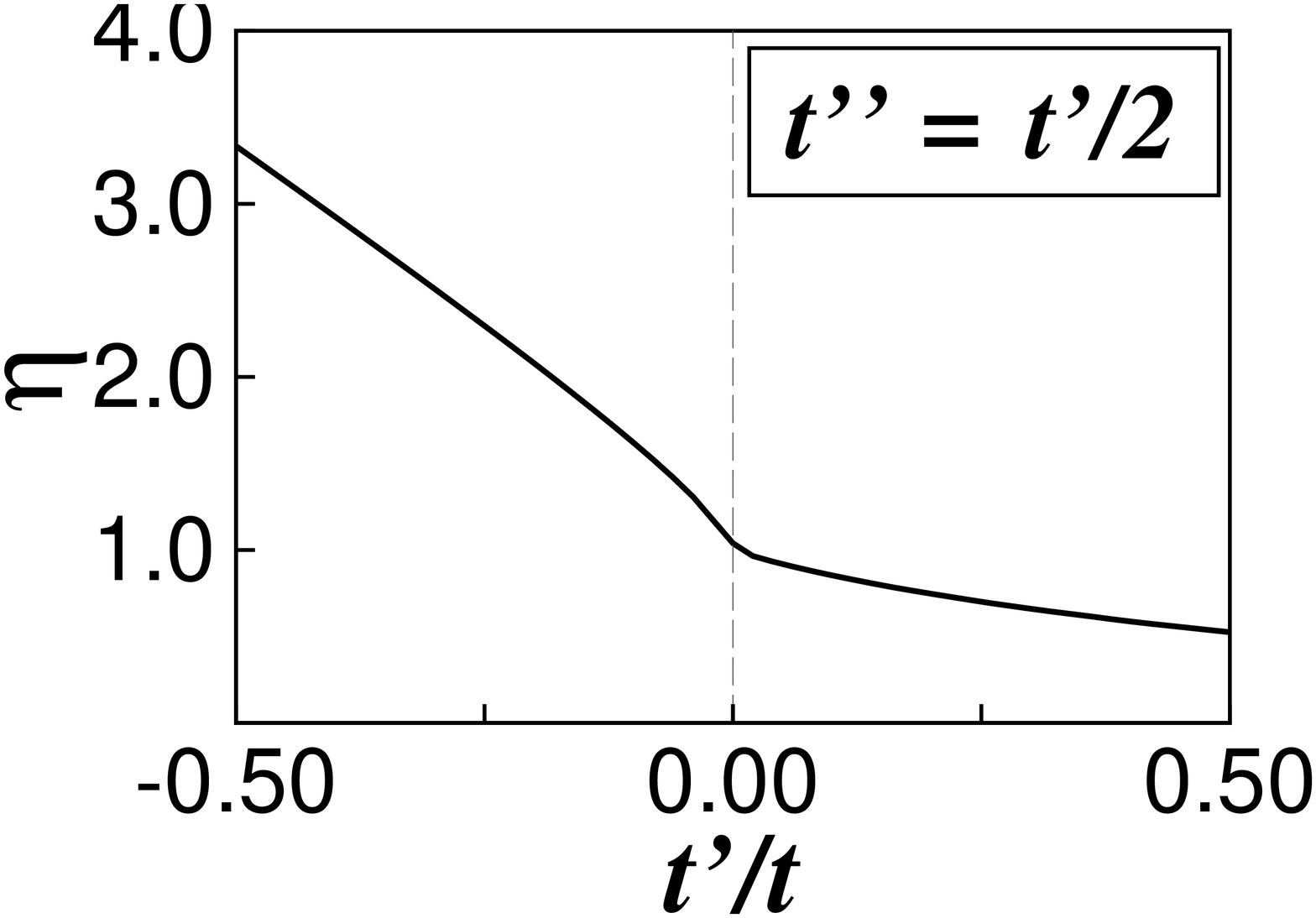}
%\centerline{\epsfysize=0.8*\stdheight \epsfbox{fscp_contour.eps} \epsfysize=0.8*\stdheight \epsfbox{tppeqtpby2.eps}}
\caption{ (color online) Left: A contour plot of $\eta$ as a function of the hopping amplitude $t'/t$ and $t''/t$. Right: Monotonic relationship between $\eta$ and the range parameter~\cite{Pavarini2001} $t'/t$ (for the special case of $t''=t'/2$).}
\label{FSCPCont}
\vspace{-0.5cm}
\end{figure}

For the cases of our interest, the ``Fermi surface" at half filling can be described as a curve in the first Brillouin zone
$k_F(\theta)$ (in polar coordinates) with $\theta$ measured from the $k_x$ axis. We define
%\bea 
%\eta\left(t'/t,t''/t \right)  = 2\left[\frac{k_F(\pi/4)}{k_F(\theta_{\rm{min}})}\right]^2 ,
%\label{FSCP} 
%\eea 
\bea 
\eta\left(t'/t,t''/t \right)  = 2\left[k_F(\pi/4) / k_F(\theta_{\rm{min}}) \right]^2 ,
\label{FSCP} 
\eea 
where $\theta_{\rm{min}}$ is the minimum angle at which a Fermi crossing exists.
In the electron-doped case (Fig.\ref{PhDiag}(b)), $\theta_{min}=0$ and $\eta>1$ corresponding to a convex 
``Fermi surface".
In the hole-doped case (Fig.\ref{PhDiag}(c)), $\theta_{min}$ corresponds to the Fermi crossing
on the zone boundary leading to a concave ``Fermi surface" with $\eta < 1$.
For $t'=t''=0$, the particle-hole symmetric case, $\eta=1$.
In Fig.~\ref{FSCPCont} we show the dependence of $\eta$ on
$t'$ and $t''$. The right panel focusses on the $t'$-dependence of $\eta$ on the special line $t''=t'/2$ which corresponds to the parameters obtained by Pavarini {\it et al.}~\cite{Pavarini2001} from electronic structure calculations of single layer cuprates.

\begin{figure}
\includegraphics[width=2.0in,angle=0]{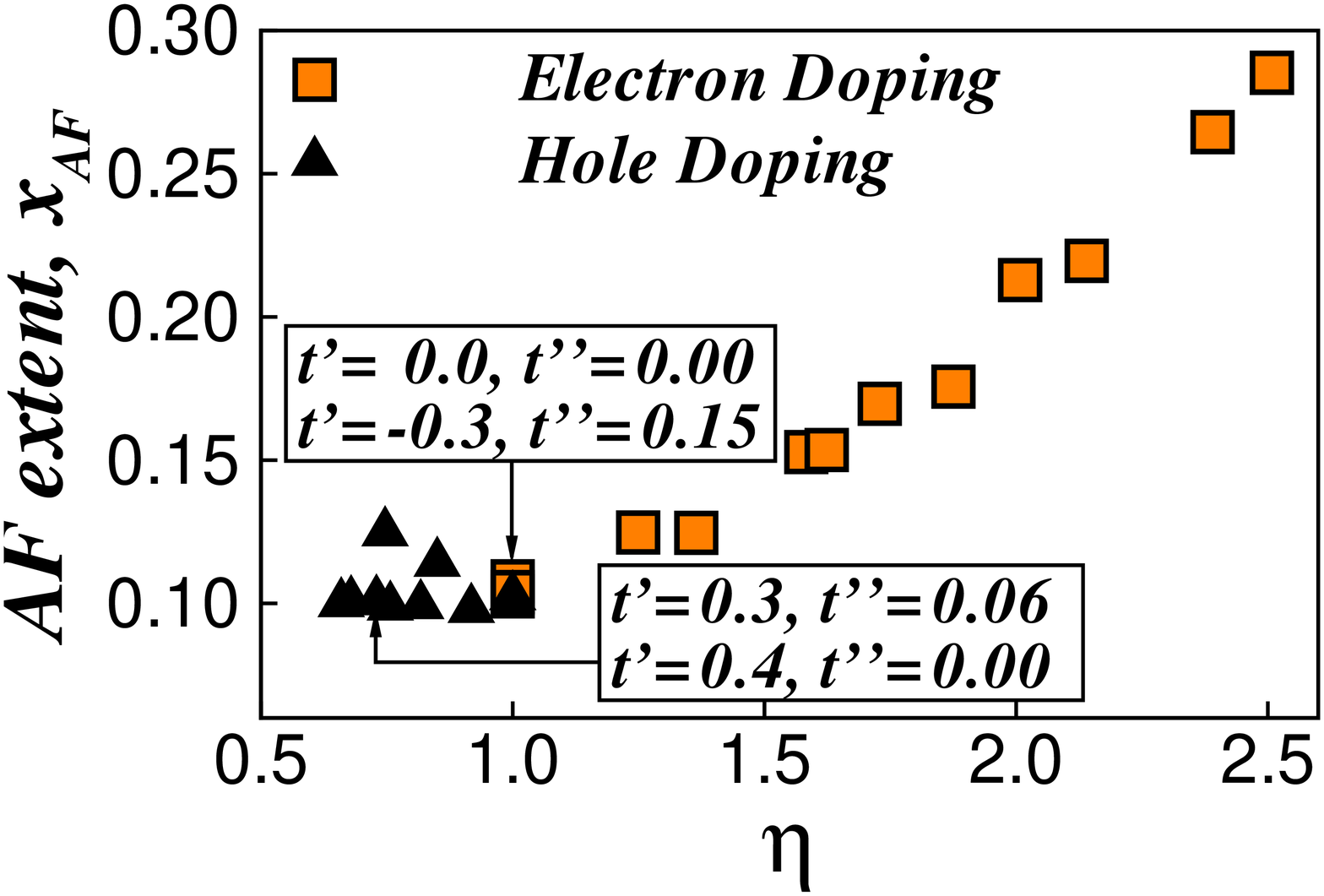}
\includegraphics[width=2.0in,angle=0]{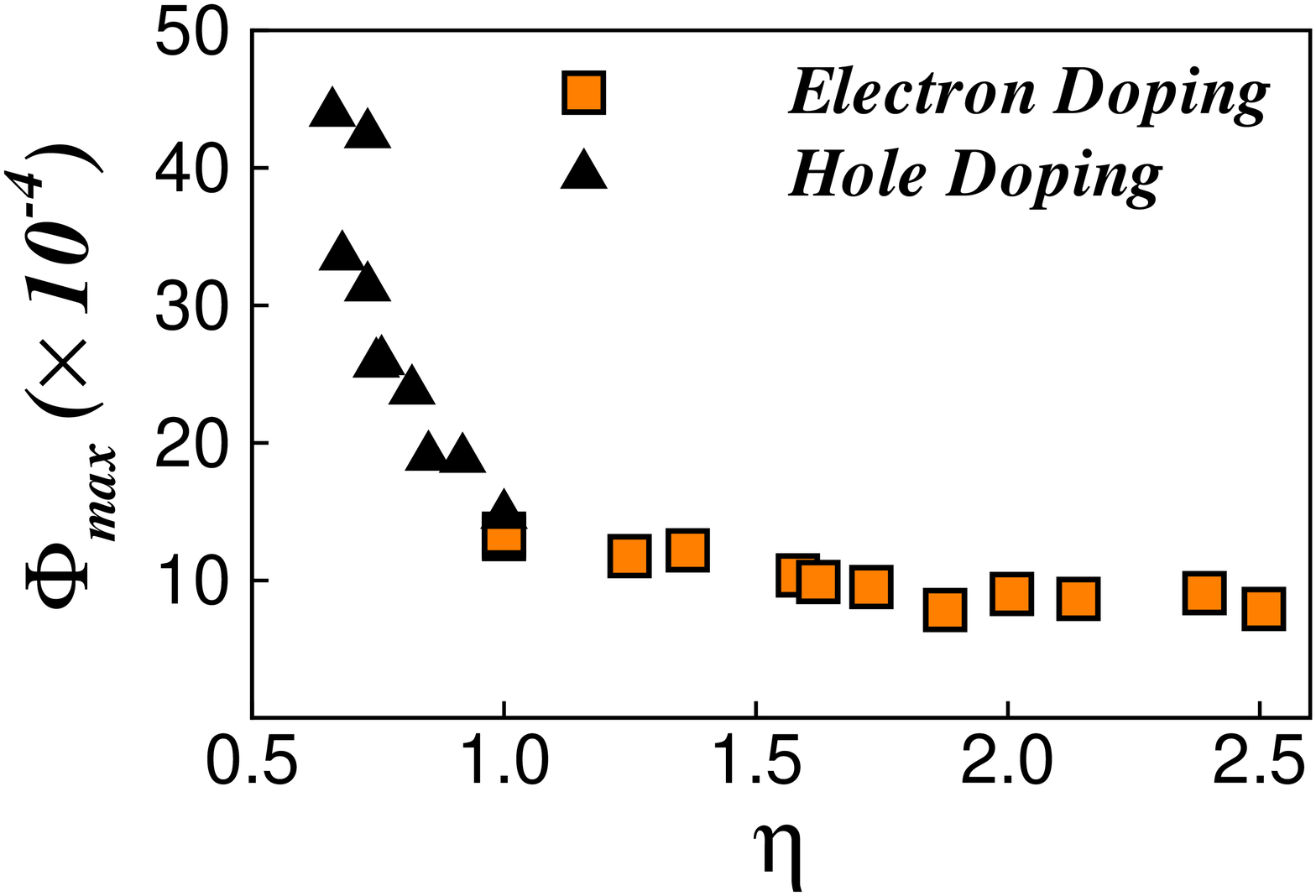}
%\centerline{\epsfysize=\stdheight \epsfbox{AFCritDopvskeop.eps}  \epsfysize=\stdheight \epsfbox{odlrovskeop.eps}}
  \caption{ (color online) Top: The size of AF phase $x_{AF}$ as a function of $\eta$. 
  Note that systems with very different bare $t$s but with the {\it same} $\eta$ have similar $x_{AF}$. 
  This is also true for $\Phi_{max}$ and optimal doping $x_{\rm opt}$. 
  Bottom: SC order at optimal doping  $\Phi_{max}$ as a function of $\eta$.}
  \label{pramvFSCP}
  \vspace{-0.5cm}
\end{figure}

In \Fig{pramvFSCP} we show the $\eta$-dependence of the 
critical doping $x_{AF}$ for the vanishing of AF
order. On the electron doped side, $x_{AF}$
increases approximately linearly with $\eta$, while for hole doping
$x_{AF}$ is essentially independent of $\eta$. 
Turning now to superconductivity, we see from
Fig.~\ref{pramvFSCP} that the SC order parameter $\Phi_{max}$ at optimality increases roughly linearly with
decreasing $\eta$ on the hole doped side. On the electron doped side,
there is a slight linear decrease of $\Phi_{max}$ as a function of
increasing $\eta$. Thus a more concave bare ``Fermi surface" leads to a more stable
SC state.  
We have also studied the dependence of the
``optimal doping'' $x_{\rm {opt}}$ (doping at which $\Phi$ attains
$\Phi_{max}$) and found a similar correlation with $\eta$. 

We have thus demonstrated that three characteristics of the phase diagram, 
$x_{AF}, \Phi_{max}$ and $x_{\rm opt}$, are determined by a single parameter 
$\eta$, rather than by the details of the bare dispersion. Two systems with a given $t$ and $J$
but with two very different $t',t''$ have the same phase diagram provided they correspond to the same value of $\eta$.
This point is clearly illustrated by the specially marked points in Fig.~\ref{pramvFSCP}.

\begin{figure}
% \centerline{\epsfxsize=12.0truecm \epsfbox{kevsx_various_fscp.eps}}
% \epsfxsize=\halfwidefig \epsfbox{kevsx_various_fscp.eps}
\centerline{$\!\!\!\!\!\!\!\!\!\!$ \epsfxsize=1.7in \epsfbox{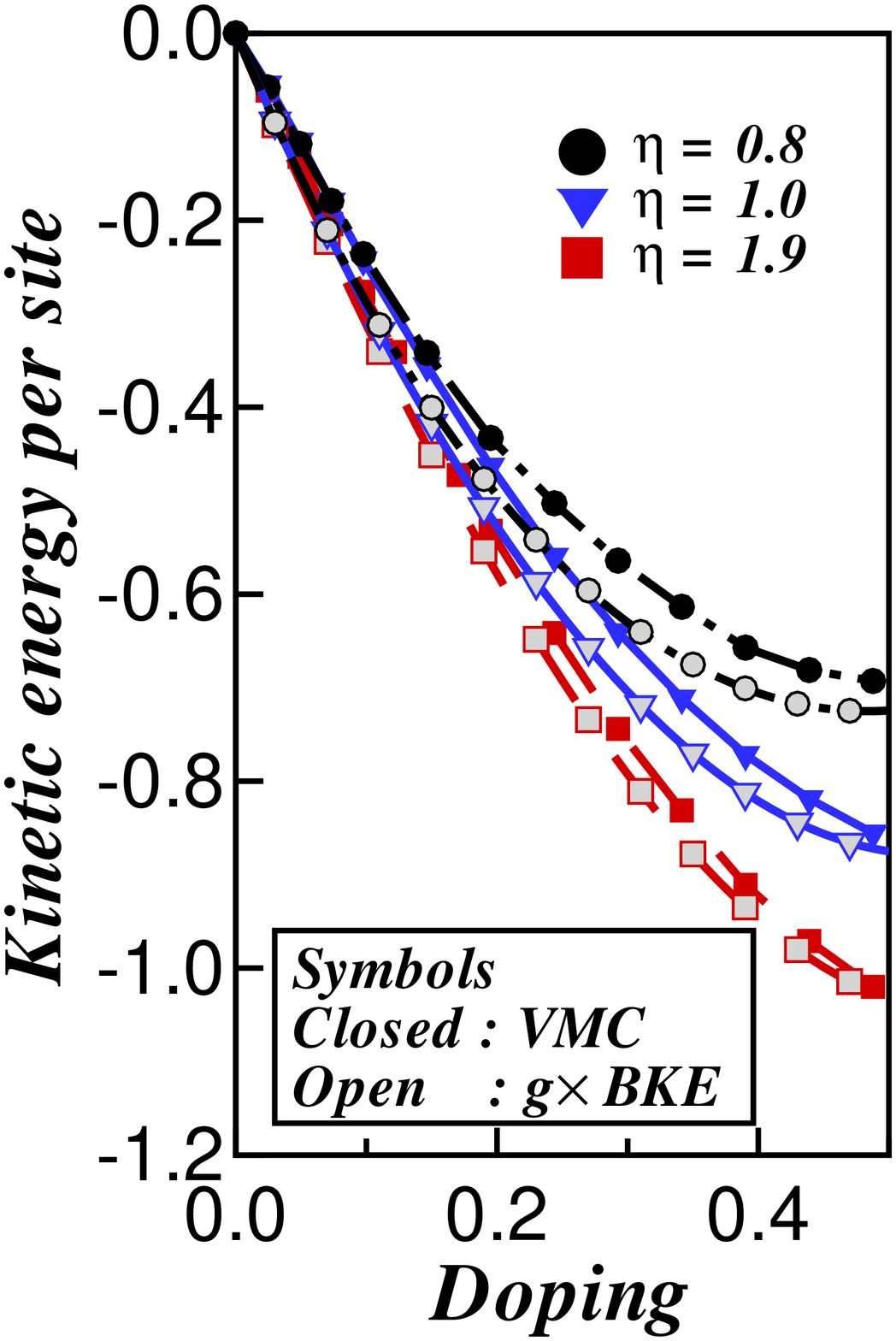} \epsfxsize=1.7in \epsfbox{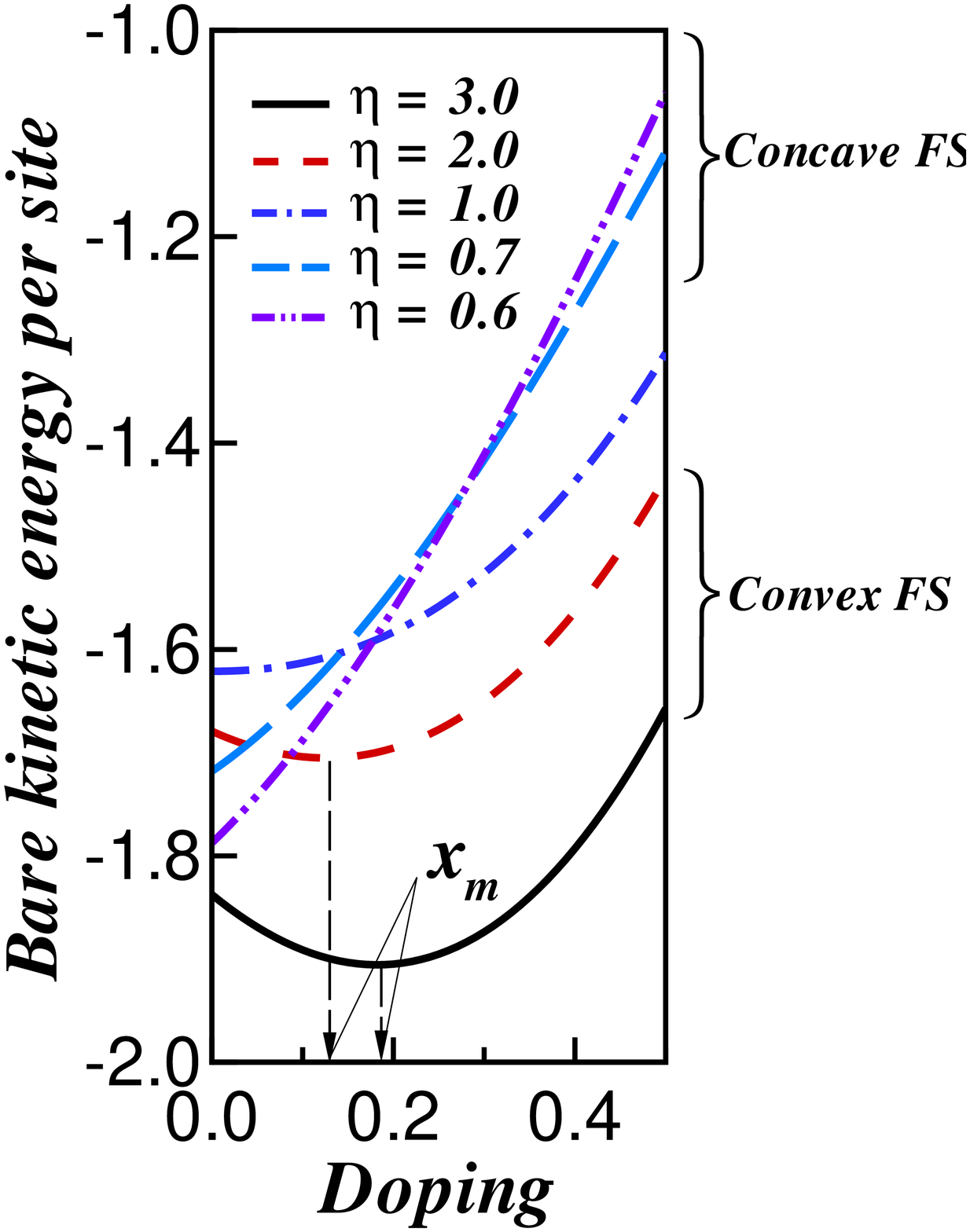}}
  \caption{ (color online) Left: Comparison of kinetic energy per site obtained from variational Monte Carlo and Gutzwiller projected bare kinetic energy (BKE); ($g = {2x}/{(1+x})$) Right: Bare kinetic energy as a function of doping for systems with different $\eta$. }
  \label{KEplots}
  \vspace{-0.5cm}
\end{figure}

To understand how the single parameter $\eta$ controls the entire
phase diagram, we consider the competition between the kinetic and exchange energies.  
Upon doping the Mott insulator,
the holes (or electrons) attempt to gain kinetic energy (KE).
Near half filling, the hopping $t$ between different sublattices
disrupts the anti-ferromagnetic order and increases the exchange
energy.  On the other hand, the hoppings $t'$ and $t''$ between the
same sublattice do not disturb AF order and entail no exchange energy
penalty.  Insight into how the system gains the most KE, while keeping
the exchange energy increase to a minimum, may be obtained from
studying the behavior of the KE as a function of doping $x$.

In Fig.~\ref{KEplots}(a), we show that the variational Monte Carlo
(projected) KE for a convex ($\eta > 1$), a concave
($\eta < 1$), and a half-filled diamond ($\eta = 1$) Fermi surface.
In each case the projected KE is closely reproduced by the Gutzwiller
approximation \cite{Anderson2004} result $g \, {\rm K}_{\rm bare}$
where ${\rm K}_{\rm bare}$ is the bare KE in the unprojected state,
and $g = 2x/(1+x)$ the renormalization factor that takes into account
projection.

The renormalization factor $g$ is independent of the $\eta$-parameter,
and thus we focus on the \emph{bare} KE in Fig.~\ref{KEplots}(b) to
understand the $\eta$-dependence of the competition between KE and
superexchange.  For electron doping with $\eta > 1$, the bare KE is a
non-monotonic function of doping with a minimum at a finite
$x_m$. Thus up to a doping of $x_m$, one can gain KE due to $t'$ and
$t''$ without sacrificing exchange energy, thereby stabilizing the AF
state. The doping $x_m$ increases with increasing $\eta$ (see
Fig.~\ref{KEplots}(b)) which underlies the $\eta$-dependence of
$x_{AF}$ on the electron doped side.  For the p-h symmetric
($t'=t''=0$) case, $\eta =1$ and the minimum KE is at $x_m=0$.  For
hole doping with $\eta < 1$, the bare KE increases monotonically with
doping.  The exchange energy is best satisfied by means of SC with
resonating singlet pairs, while providing for the necessary KE gain,
respecting the no double occupancy constraint. Thus a system with a
larger bare KE favors a more stable SC state.

As noted earlier, Pavarini {\it et al.}~\cite{Pavarini2001} suggested
an empirical correlation between the range parameter related to $t'/t$
(for the special case of $t'' = |t'|/2$) and $T_c^{\rm max}$, the
maximum $T_c$ within a given cuprate family.  We cannot, of course,
obtain $T_c$ from our ground state calculation, but the magnitude of
the SC order parameter $\Phi$ can be
taken~\cite{Paramekanti2004,Shih2004} as a measure of $T_c$.  The
strong linear correlation between $\Phi^{max}$ and $\eta$ on the
hole-doped side seen in Fig.~\ref{pramvFSCP} is equivalent to a
similar correlation between $T_c^{\rm max}$ and the range parameter.
(Note the relationship between $\eta$ and the range parameter in
Fig.~\ref{FSCPCont}).  On the electron-doped side, however, we do not
predict a strong dependence of $T_c^{\rm max}$ on the $\eta$
parameter.

In conclusion, our results provide a unified microscopic picture for
the competition between AF and SC for electron and hole doped
cuprates, while also providing insights into a key material parameter
$\eta$ that controls many aspects of the phase diagram, including
$T_c^{\rm max}$.  In particular, this work suggests a route to
increase $T_c^{max}$ by creating systems with a small $\eta$, i.~e.,
with a highly concave bare ``Fermi surface" at half filing.  Even if
such band structure engineering may not be easy in solid state
materials, it may be possible in optical lattice realizations of
strongly correlated Fermions.

We thank DST, India for support through a SERC project (SP, VBS) and the Ramanujan grant (VBS). 
We acknowledge support from NSF-DMR 0706203 (MR) and DOE DE-FG02-07ER46423 (NT). 
We thank T.~V.~Ramakrishnan, H.~R.~Krishnamurthy, 
T.~Saha-Dasgupta, and R. Sensarma for stimulating discussions.
We acknowledge the use of computational facilities of S.~Ramaswamy and P.~Maiti
and the Ohio Supercomputer Center.

\end{document}